\documentclass[twocolumn,showpacs,preprintnumbers,amsmath,amssymb]{revtex4}
\usepackage{graphicx}
\usepackage{wasysym}
\begin{document}

\title{Scaling of fluctuations in a colloidal glass}
\author{P. Wang, C. Song and H. A. Makse}

\affiliation { Levich Institute and Physics Department
\\ City College of New York
\\ New York, NY 10031, US}

\begin{abstract}

We report experimental measurements of particle dynamics in a
colloidal glass in order to understand the dynamical
heterogeneities associated with the cooperative motion of the
particles in the glassy regime. We study the local and global
fluctuation of correlation and response functions in an aging
colloidal glass. The observables display universal scaling
behavior following a modified power-law, with a plateau dominating
the less heterogeneous short-time regime and a power-law tail
dominating the highly heterogeneous long-time regime.
\end{abstract}

\maketitle

\section{ Introduction}
Mostly due to the enormous practical importance of glassy systems
there has been a vast literature describing different theoretical
frameworks for glasses, yet without a common theory applicable to
the diverse range of systems undergoing a glass transition.
Increasing the volume fraction of a colloidal system slows down
the Brownian dynamics of its constitutive particles, implying a
limiting density, $\phi_g$, above which the system can no longer
be equilibrated with its bath \cite{Pusey}. Hence, the thermal
system falls out of equilibrium on the time scale of the
experiment and thus undergoes a {\it glass transition}
\cite{Cugliandolo}. Even above $\phi_g$ the particles continue to
relax, but the nature of the relaxation is very different to that
in equilibrium. This phenomenon of a structural slow evolution
beyond the glassy state is known as ``aging'' \cite{Struik}. The
system is no longer stationary and the relaxation time is found to
increase with the age of the system, $t_w$, as measured from the
time of sample preparation.

This picture applies not only to structural glasses such as
colloids, silica and polymer melts, but also to spin-glasses,
ferromagnetic coarsening, elastic manifolds in quenched disorder
and jammed matter such as grains and emulsions
\cite{Cugliandolo,Kurchan_prl,Bouchaud}.

In order to investigate the dynamical properties of the aging
regime we consider the evolution of the correlation function
$C(t,t_w)$ and integrated response to external fields
$\chi(t,t_w)$. These quantities are separated into a stationary
(short time) and an aging part (long time): $C(t,t_w)=C_{\rm
st}(t-t_w)+C_{\rm ag}(t,t_w)$ and $\chi(t,t_w)=\chi_{\rm
st}(t-t_w)+\chi_{\rm ag}(t,t_w)$, where we have included the
explicit dependence on $t_w$ in the aging part, and $t$ is the
observation time.

We use a simple system
undergoing a glass transition: a
colloidal glass of micrometer size particles, where the
interactions between particles can be approximated as hard core
potentials \cite{Pusey,Kegel,Weeks_sci}. The system is index
matched to allow the visualization of tracer particles in the
microscope \cite{Weeks_sci}.
Owing to the simplicity of the system, we are able to follow the
trajectories of magnetic tracers embedded in the colloidal sample
and use this information
to understand the scaling of the local and global fluctuations
of the correlation functions.

\section { Experimental setup}

Our experiments use a
colloidal suspension consisting of a mixture of
poly-(methylmethacrylate) (PMMA) sterically stabilized colloidal
particles (radius $a_{p}=1.5\ \mu m$, density $\rho_{p}=1.19\
g/cm^{3}$, polydispersity $\sim 14\%$) plus a small fraction of
superparamagnetic beads (radius $a_{m}=1.6\ \mu m$ and density
$\rho_{m}=1.3\ g/cm^{3}$, from Dynal Biotech Inc.) as the tracers

The colloidal
suspension is immersed in a solution  containing $76\%$ weight
fraction of cyclohexylbromide and $24\%$ cis-decalin which are
chosen for their density and index of refraction matching
capabilities \cite{Weeks_sci}. For such a system the glass
transition occurs at
$\phi_{g}\approx 0.57 - 0.58$ \cite{Pusey,Kegel,Weeks_sci}.
In our experiments we consider two samples at different densities
and determine the glassy phase for the samples that display aging.
The main results are obtained for sample A just above the glass
transition $\phi_{\rm A}= 0.58\pm0.01$. We also consider a denser
sample B with $\phi_{\rm B}=0.60\pm0.01$, although this sample is
so deep in the glassy phase that we are not able to study the slow
relaxation of the system and the dependence of the waiting time
within the time scales of our experiments.

We use a magnetic force as the external perturbation to generate
two-dimensional motion of the tracers \cite{Song}
on a microscope stage following a simplified
design of \cite{magnetic_stage}. Video microscopy and computerized
image analysis are used to locate the tracers in each image. We
calculate the response and correlation functions in the x-y plane.

\begin{figure}
\centering \resizebox{8cm}{!}{\includegraphics{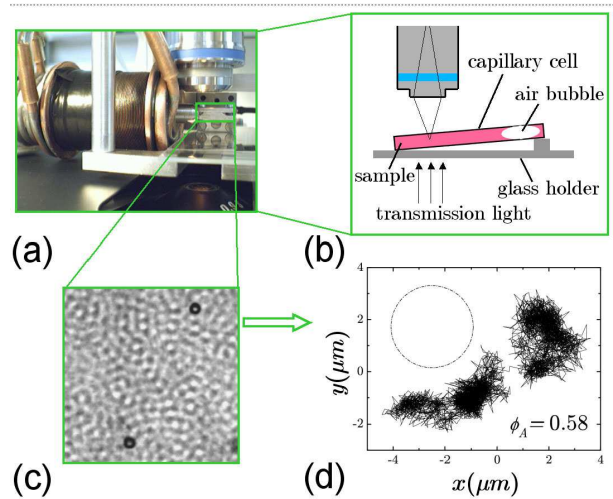}}
\caption{(a) Picture of the experimental setup. (b) Schematic
illustration of the setup. (c) Detail of an image of tracer in
sample A. (d) Trajectory of a tracer in sample A showing the cage
dynamics over 4 hours. The circle represents the size of the
tracer particle, $3.2\mu m$.} \label{images}
\end{figure}

The experimental set up is shown in Fig. \ref{images}a and Fig.
\ref{images}b. We use a Zeiss microscope with a 50$\times$
objective of numerical aperture 0.5 and a working distance of 5mm.
We work with a field of view $194\mu m\times155\mu m$ by using a
digital camera with the resolution of 1288 pixels $\times$ 1032
pixels. We locate the center of each bead position with sub-pixel
accuracy, by using image analysis. For the condensed samples A and
B, we use the low frame rate $1/3$ frame/sec, while for the dilute
sample C, we record the images at $1$ frame/sec. The long working
distance of the objective is necessary to allow the pole of the
magnet to reach a position near the sample. An example of the
images of the tracers obtained in sample A is shown in Fig.
\ref{images}c where we can see the black magnetic tracers embedded
in the background of nearly transparent PMMA particles. An example
of the trajectory in the x-y plane of a tracer diffusing without
magnetic field is shown in Fig. \ref{images}d. We note that this
particular tracer moves away from two cages in a time of the order
of $4$ hours.

The magnetic field is produced by one coil made of 1200 turns of
copper wire. We arrange the pole of the coil perpendicular to the
vertical optical axis, and generate a field with no vertical
component. Thus, the tracers move in the x-y plane with a slight
vertical motion which is generated by a density mismatch between
the tracers and the background PMMA particles. This vertical
motion is very small at the high volume fraction of interest here,
and therefore we calculate all the observables in the x-y plane.

The magnetic force is calibrated for a given coil current by
replacing the suspension with a mixture of $50:50$ water-glycerol
solution with a few magnetic tracers. The distance between the top
of the magnetic pole and the vertical optical axis is always
fixed, which means that the magnetic force at the local plane
depends only on the coil current. At a given current, we determine
the velocity of the magnetic tracers at the focal point and
calculate the magnetic force from Stokes's law, $F = 6\pi\eta
a_{m}v$, where $\eta$ is the viscosity of the water-glycerol
solution, $a_{m}$ is the tracer radius, and $v$ is the observed
velocity of the tracer. The uncertainty in the obtained force
comes from: (a) the uncertainty in the coil current which is
$1\%$, (b) the beads, which are not completely monodisperse in
their magnetic properties, causes a $10\%$ uncertainty
\cite{Weeks_force}, and (c) the magnetic field is slowly decaying
in the field of view, causing a $4\%$ uncertainty.

In order to investigate the dynamical properties of the aging
regime, we first consider the autocorrelation function as the mean
square displacement (MSD) averaged over 82 tracer particles,
$C(t,t_w) \equiv \langle\Delta x^{2}(t,
t_w)\rangle/2=\langle[x(t_w+\Delta t, t_w)-x(t_w,
t_w)]^2\rangle/2$, at a given observation time, $t=t_w+\Delta t$,
after the sample has been aging for $t_w$ as measured from the end
of the stirring process. Then, we measure the integrated response
function (by adding the external magnetic force, $F$) given by the
average position of the tracers, $\chi(t,t_w) \equiv \langle x(t_w
+\Delta t, t_w ) - x(t_w ,t_w ) \rangle /F$.


\section{Sample Preparation at $t_w=0$}
\label{inistate}

\begin{figure}
\centering \resizebox{8cm}{!}{\includegraphics{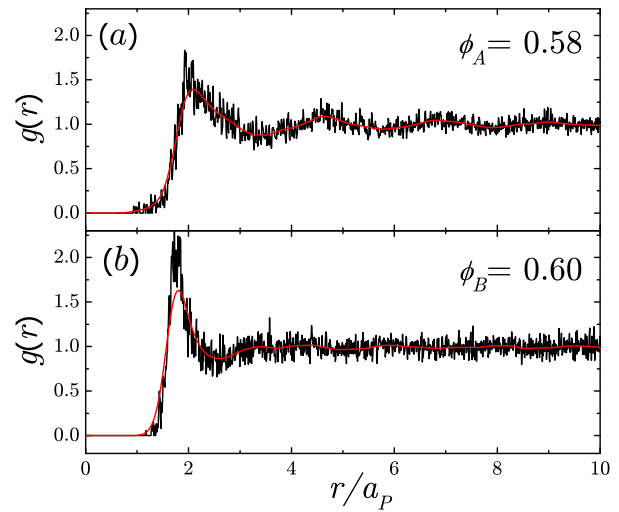}}
\caption{Pair-distribution functions of (a) sample A and (b)
sample B.}\label{gr}
\end{figure}

It is important to correctly determine the initial time of each
measurement to reproduce the subsequent particle dynamics. We
initialize the system by stirring the sample for two hours with an
air bubble inside the sample (see Fig. \ref{images}b) to
homogenize the whole system and break up any pre-existing
crystalline regions. Then we place the sample on the magnetic
stage and take images to obtain the trajectories of the tracers
that appear in the field of view. The initial times $t = t_w = 0$
are defined at the end of the stirring. After measuring all the
tracers appearing in the field of view, a new stirring is applied,
the waiting time is reset to zero, and the measurements are
repeated for a new set of tracers. We have analyzed the
pair-distribution function, $g(r)$, and two-time intensity
autocorrelation function, $g_{2}(t_w,t_w+ \Delta t)$, in order to
test our rejuvenation technique.

First, Fig. \ref{gr} plots the pair-distribution functions of
sample A and sample B right after the stirring procedure. We
calculate the pair-distribution functions by reconstructing the
packings from 3D confocal microscopy images of size $60\mu
m{\times}60\mu m{\times}15\mu m$. We find that the samples do not
show obvious crystallized region by directly looking either at the
pair-distribution function or the images taken from confocal
microscopy.

For these measurements we use a Leica confocal microscope. The
PMMA particles are fluorescently dyed so that they are ready to be
observed by confocal microscopy. We load the samples sealed in a
glass cell on the confocal microscope stage and use a  Leica HCX
PL APO  63x, 1.40 numerical aperture, oil immersion lens for 3D
particle visualization in order to calculate the volume fraction
and the pair-distribution function.

\begin{figure}
\centering \resizebox{8cm}{!}{\includegraphics{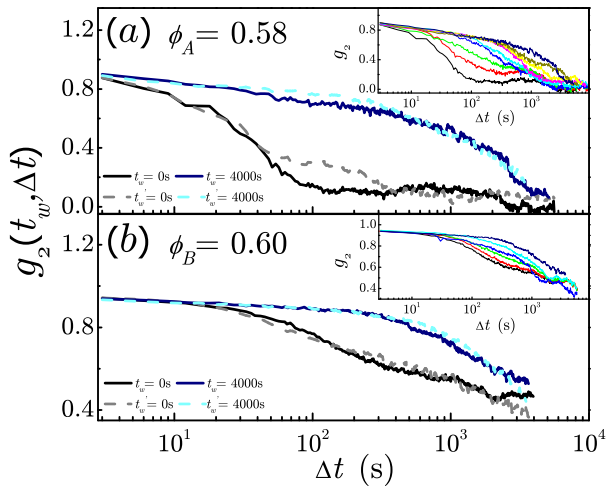}}
\caption{Two-time intensity autocorrelation function, $g_2(t_w+
\Delta t, t_w)$, of (a) sample A and (b) sample B. The solid black
and navy curves correspond to $t_w=0$s and $t_w=4000$s,
respectively. The dash gray and light blue curves correspond to
the measurements after the second stirring with ${t_w}^{'}=0$s and
${t_w}^{'}=4000$s, respectively. The fact that the curves at
$t_w=0$s and ${t_w}^{'}=0$s coincide indicates that sample has
been fully rejuvenated to its initial state. The inset is the same
plot as the main figure for different $t_w$ ranging from 0s (solid
black) to 4000s (solid navy). }\label{g2}
\end{figure}

Second, we analyze $g_2(t_w+ \Delta t, t_w)$ and show that the
sample has been rejuvenated by the stirring process. We plot
$g_2(t_w+ \Delta t, t_w)$ for 3 situations: (a) after the first
stirring process ($t_w=0$), (b) after the sample has aged for a
long time ($t_w=4000s$), (c) we then apply the stirring again and
immediately plot the autocorrelation function for the new initial
time (${t_w}^{'}=0$). The plot show the lack of correlation for
the two individual measurements $t_w=0$ and ${t_w}^{'}=0$, thus
demonstrating that the sample has been rejuvenated. Technically,
we record the temporal image sequence of sample A and sample B,
and study the aging by calculating two images' correlation defined
as:

\begin{equation}
g_{2}(t_w,t_w+ \Delta t)=\frac{\langle I(t_w)I(t_w+\Delta
t)\rangle}{\langle I(t_w)^2\rangle},
\end{equation}
where $I$ is the average gray-scale intensity of a small box (20
pixels $\times$ 20 pixels, PMMA particles' diameter is roughly
equal to 20 pixels), and $I(t_w)$ and $I(t)$ come from the images
at different times $t_w$ and $t$, respectively. We cut one image
(1288 pixels $\times$ 1032 pixels) into many boxes, and calculate
the correlation of two boxes located at the same position in the
two images. The average $\langle...\rangle$ is taken over all the
boxes in one image. Fig. \ref{g2} plots the correlation function,
$g_{2}(t_w,t_w+ \Delta t)$, of sample A and sample B, which shows
aging behavior (see the inset). Fig. \ref{g2} shows how
$g_{2}(t_w,t_w+ \Delta t)$ calculated after two stirring processes
at $t_w=0$ and ${t_w}^{'}=0$ coincide, indicating that the sample
can be fully rejuvenated to its initial state by our stirring
technique.

\begin{figure}
\centering \resizebox{8cm}{!}{\includegraphics{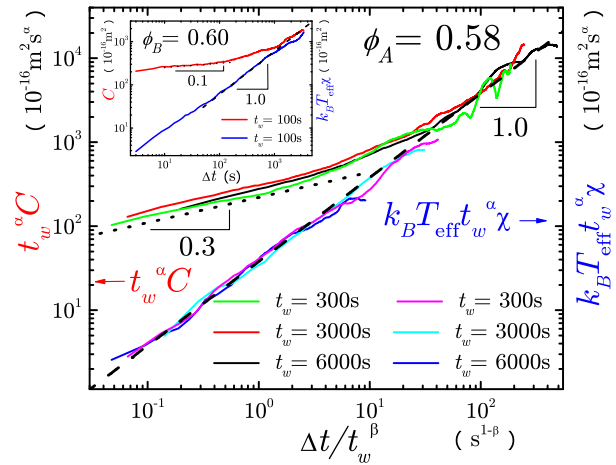}}
\caption{Scaling plot of sample A for the scaled autocorrelation,
${t_w }^{\alpha}C$, and scaled integrated response, $k_{B}T_{\rm
eff}{t_w }^{\alpha}\chi$, as a function of the time ratio,
$\Delta{t}/{t_w }^{\beta}$, for different waiting times. The black
dash line is a linear fit which indicates that $T_{\rm eff}=690K$.
The inset is a plot of sample B for autocorrelation, $C$, and
integrated response, $k_{B}T_{\rm eff}\chi$, as a function of
$\Delta{t}$ at $t_{w}=100s$. The black dash line is a linear fit
which indicates that $T_{\rm eff}=1600K$.} \label{scaling}
\end{figure}

\begin{figure}
\centering \resizebox{6cm}{!}{\includegraphics{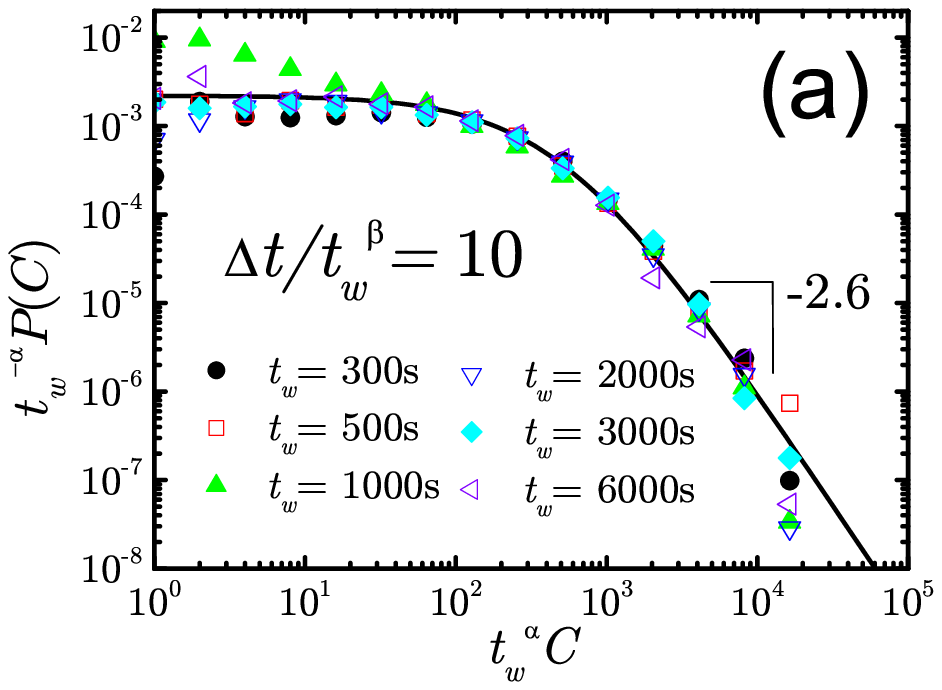}}
\centering \resizebox{6cm}{!}{\includegraphics{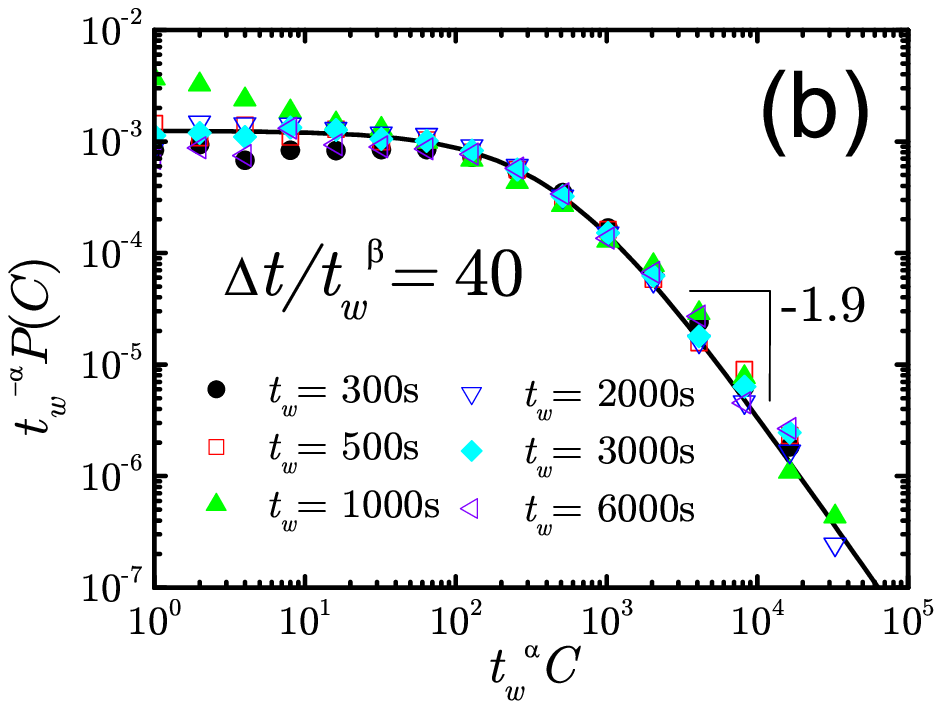}}
\centering \resizebox{6cm}{!}{\includegraphics{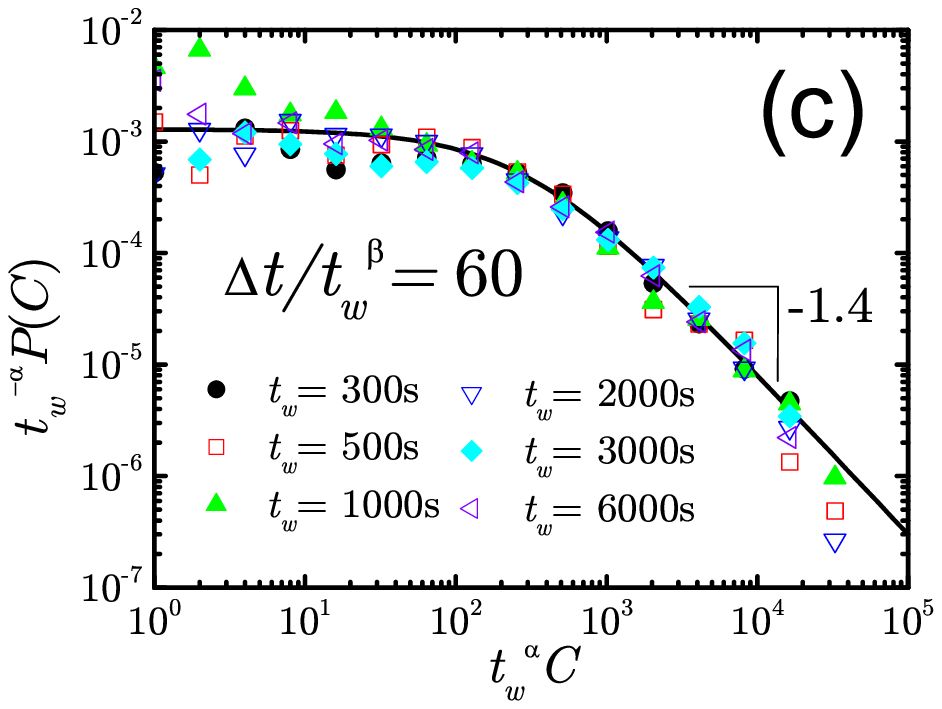}}
\centering \resizebox{6cm}{!}{\includegraphics{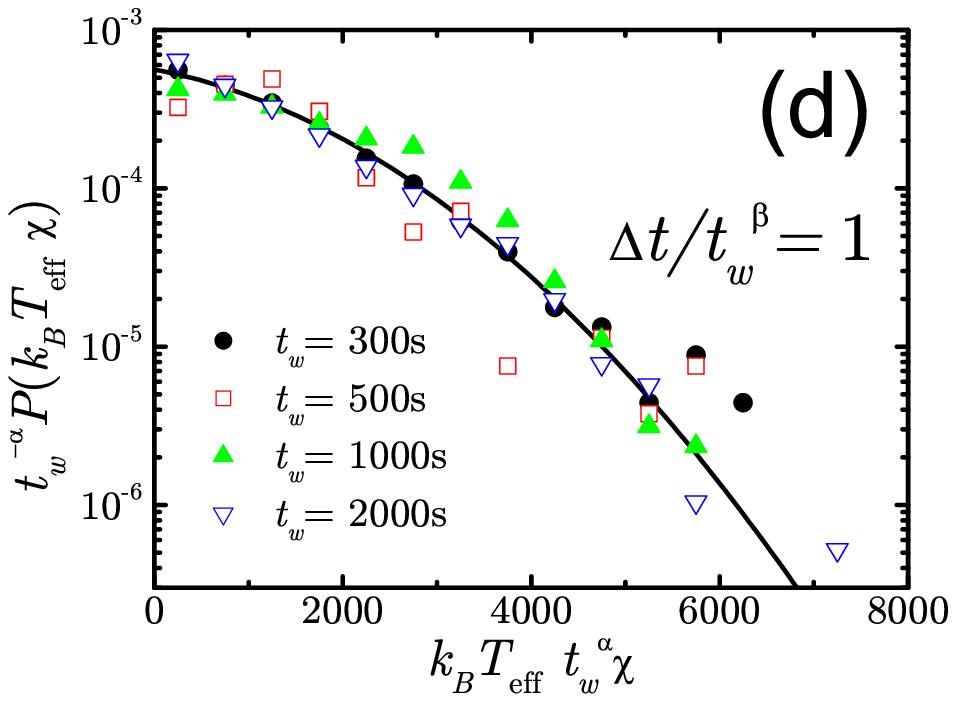}}
\caption{(a) Scaled $\langle x_{12}\rangle$ as a function of
scaled $r_{01}$ for given ratio $\Delta t/t_{w}^{\beta}=10$. The
linear part for different $t_{w}$ and $\Delta t$ are coincident
and well described as $t_{w}^{\alpha/2}\langle
x12\rangle=(-0.28)t_{w}^{\alpha/2}r_{01}$ , indicated by the
dashed line. (b) Probability distribution function(PDF) of the
scaled local correlations for the ratios $\Delta
t/t_{w}^{\beta}=1$. (c) Probability distribution function(PDF) of
the scaled local correlations for the ratios $\Delta
t/t_{w}^{\beta}=60$. The inset is a log-log plot of averaged PDF
of the scaled local correlations for three ratios $\Delta
t/t_{w}^{\beta}=1, 10, 60$, the tails are coincident and well
fitted by power law with a index -1.25. (d) Probability
distribution function(PDF) of the scaled local integrated
responses for the ratios $\Delta t/t_{w}^{\beta}=1$. The inset is
a log-linear plot of averaged PDF of the scaled local integrated
responses for the ratios $\Delta t/t_{w}^{\beta}=1$, the tail are
well fitted by a Gaussian function
$exp(0.955+0.00881x-5.69613x^{2})$. } \label{pdf}
\end{figure}

\section{Scaling ansatz for the global correlations and responses.}

Insight into the understanding of the slow relaxation can be
obtained from the study of the universal dynamic scaling of the
observables with $t_w$. Based on spin-glass models, different
scaling scenarios have been proposed \cite{Cugliandolo,Henkel} for
correlation and response functions. Our analysis indicates that
the observables can be described as

\begin{equation}
\begin{array}{ll}
C(t_w +\Delta t, t_w )= \langle \Delta x^{2} \rangle/2 =
t_w^{-\alpha}f_{D}(\frac{\Delta t}{{t_w }^{\beta}}), \\
\chi(t_w+\Delta t, t_w )= \frac{\langle \Delta x \rangle}{F} =
t_w^{-\alpha}f_{M}(\frac{\Delta t}{{t_w }^{\beta}}),
\end{array}
\label{scaling-laws}
\end{equation}

where $f_{D}$ and $f_{M}$ are two universal functions and $\alpha$
and $\beta$ are the aging exponents. Evidence for the validity of
these scaling laws is provided in Fig. \ref{scaling} where the
data of the correlation function and the integrated response
function collapse onto a master curve when plotted as $
t_w^{\alpha}C(t_w +\Delta t,t_w )$ and $t_w^{\alpha}\chi(t_w
+\Delta t,t_w )$ versus $\Delta t/{t_w }^{\beta}$. By minimizing
the $\sigma^2$ value of the difference between the master curve
and the data we find that the best data collapse is obtained for
the following aging exponents: $\alpha + \beta = 0.34\pm0.05$ and
$\beta = 0.48\pm0.05$. We find (Fig. \ref{scaling}) that the
scaling functions satisfy the following asymptotic behavior:
\begin{equation}
f_{D}(y)\sim
f_{M}(y)\sim y,
\end{equation}
in agreement with the fact that the motion of the particles is
diffusive at long times,
 $\langle \Delta x^{2} \rangle \sim \Delta t$,
and the existence of a well-defined mobility, respectively.
Therefore at long times, both the correlation and response
functions display the same power law decay:
\begin{equation}
\begin{array}{ll}
C(t_w +\Delta t,t_w ) \sim  t_w^{-(\alpha+\beta)} \Delta t, \\
\chi(t_w +\Delta t,t_w ) \sim  t_w^{-(\alpha+\beta)} \Delta t.
\end{array}
\end{equation}

The result $C(t_w +\Delta t, t_w) \sim \chi(t_w+\Delta t, t_w)
\sim t_w^{-0.34}$ indicates
$D(t_w )\sim M(t_w) \sim t_w ^{-0.32}$.
For short times, the MSD scaling
function crosses over to a sub-diffusive behavior of the
particles. We obtain,  $\langle x^2(t_w +\Delta t,t_w )\rangle
\sim t_w^{-\alpha}(\Delta t/t_w^{\beta})^{0.3}=t_w^{-0.004}\Delta
t^{0.3}$. Since the trapping time corresponds to the size of the
cages denoted by $q(t_w)$, we can determine the cage dependence on
$t_w$ as $q(t_w)\sim t_w^{-0.002}$. The resulting exponent is so
small that we can say that the cages are not evolving with the
waiting time, within experimental uncertainty. Furthermore, the
scaling ansatz of Eq. (\ref{scaling-laws}) indicates that the
relaxation time of the cages scales as $\tau(t_w)\sim t_w^{\beta}$
since this is the time when the sub-diffusive behavior
crosses-over to the long time diffusive regime.

We find that the scaling forms of the correlations and responses
are not consistent with p-spin models. Based on invariance
properties under time reparametrization, spin-glass models predict
a general scaling form $C_{\rm ag}(t,t_w ) = C_{\rm
ag}(h(t)/h(t_w))$, where $h(t)$ is a generic monotonic function
\cite{Cugliandolo}. We find that the scaling of our observables
cannot be collapsed with the ratio $h(t)/h(t_w)$. The scaling with
$h(t)/h(t_w)$ is expected for system in which the correlation
function saturates at long times \cite{Cug_Dou}. On the other
hand, our system is diffusive, and the studied correlation
function is not bounded. Indeed, similar scaling as in our system
has been found in the aging dynamics of anther unbounded system:
an elastic manifold in a disordered media \cite{Bustingorry}. The
suggestion is that this problem and the particle diffusing in a
colloidal glass may belong to the same universality class.
Furthermore our results can be interpreted in terms of the droplet
picture of the aging of spin glass, where the growth of the
dynamical heterogeneities control the aging.

\section{ Local fluctuations of autocorrelations and responses.}

Previous work has revealed the existence of dynamical
heterogeneities, associated with the cooperative motion of the
particles, as a precursor to the glass transition as well as in
the glassy state \cite{Kegel,Weeks_sci,Weeks_aging,Kob}. Instead
of the average global quantities studied above, the existence of
dynamical heterogeneities requires a microscopic insight into the
structure of the glassy. Earlier studies focused mainly on
probability distributions of the particles displacement near the
glass transition. More recent analytical work in spin glasses
\cite{Castillo} shows that the probability distribution function
(PDF) of the local correlation $P(C)$ and the local integrated
response $P(\chi)$ could reveal essential features of the
dynamical heterogeneities.

Here we perform a systematic study of $P(C)$ and $P(\chi)$ in
sample A, and the resulting PDFs are shown in Fig. \ref{pdf}.
The scaling ansatz of Eq. (\ref{scaling-laws}) implies that $P(C)$
and $P(\chi)$ should be collapsed by rescaling the time $\Delta t$
by $t_w^{\beta}$ and the local fluctuations by $t_w^{\alpha}$.
Indeed, this scaling ansatz provides the correct collapse of all
the local fluctuations captured by the PDFs, as shown in Fig.
\ref{pdf}a, \ref{pdf}b and \ref{pdf}c for $P(C)$ and in Fig.
\ref{pdf}d for $P(\chi)$.

The PDF of the autocorrelation function displays a universal
behavior following a modified power-law

\begin{equation}
t_w^{-\alpha}P(C)\propto(t_w^\alpha C+C_0)^{-\lambda}, \label{mpl}
\end{equation}
where
$C_0$ and $\lambda$ only depend on the time ratio $\Delta
t/t_w^\beta$. For the smaller values of $C$ ($C<t_w^{-\alpha}
C_0$), the existence of a flat plateau in $P(C)$ indicates that
the tracers are confined in the cage. For larger values of $C$,
the salient feature of the PDF is the very broad character of the
distribution, with an asymptotic behavior $P(C)\sim C^{-\lambda}$.
This large deviation from a Gaussian behavior is a clear
indication of the heterogeneous character of the dynamics.
Furthermore, the exponent $\lambda$ decreases from 2.6 to 1.4 with
the time ratio $\Delta t/t_w^{\beta}$ ranging from 10 to 60. We
notice that $\lambda =2$ corresponds to the crossover between the
short-time and long-time regime in Fig. \ref{scaling}, where
$\Delta t/t_w^\beta\approx40$. The significance of $\lambda=2$ is
seen in the integral $\int P(C)CdC$. For $\lambda>2$ ($\Delta
t/t_w^\beta < 40$) the plateau dominates over the power law tail
in the integral and the dynamics is less heterogeneous. For
$\lambda<2$ ($\Delta t/t_w^\beta > 40$) the power law tail
dominates and this regime corresponds to the highly heterogeneous
long-time regime.

On the contrary, $P(\chi)$, shown in Fig. \ref{pdf}d, displays a
different behavior. The fluctuations are more narrow and the PDF
can be approximated by a Gaussian. This is consistent with the
fact that we did not find cage dynamics for the global response.
Moreover, numerical simulations of spin-glass models
\cite{Castillo} seem to indicate a narrower distribution as found
here.

\section{Scaling behavior of local fluctuations}
\label{local}

The local correlation function $C$ and local integrated response
function $\chi$ studied here are calculated from each individual
tracer trajectory. In a condensed colloidal sample, the tracer
trajectories are always confined at a local position. Therefore,
the correlation $C$ and response $\chi$ for an individual particle
can be regarded as the coarse-grained {\it local} fluctuations of
the observables as investigated in \cite{Castillo}.

Furthermore, in order to improve the statistics of our results,
the PDF $P(C)$ is calculated not only for all the tracers, but
also over a time interval much smaller than the age of the system.
In practice, we calculate the $i-$th tracer's local correlation
function as $C_{i}(t_w+\Delta t, t_w)$. At a given $\Delta t$, we
open a small time window $[\Delta t-t_{s},\Delta t+t_{s}]$ ($t_s
\ll \Delta t$) and count all the $C_{i}(t_w+\Delta t_j, t_w)$ with
$\Delta t_j\in[\Delta t-t_{s},\Delta t+t_{s}]$ into the statistics
of $P(C)$. The calculated $P(C)$ is a mixture of the local and the
temporal fluctuations of the observables. Similar technique is
performed to calculate $P(\chi)$.

Next we derive the scaling law for the PDF of the local
correlation function. Let us first recall the scaling behavior of
the global correlation function in Eq. (\ref{scaling-laws}),

\begin{equation}\label{gC}
t_w^{\alpha}\overline{C(t_w +\Delta t,t_w )}=f_D({\Delta t}/{{t_w
}^{\beta}}),
\end{equation}
where we add a bar to $\overline{C}$ to distinguish the global
correlations from the local $C$. The average is taken over all the
tracer particles. We can rewrite the global correlation function
as the integration of the PDF $P(C)$ as:

\begin{equation}\label{lC}
\overline{C}= \int P(C)CdC.
\end{equation}

Furthermore, we obtain the relation of $f_D$ and $P(C)$ by
substituting Eq. (\ref{lC}) into Eq. (\ref{gC}):

\begin{equation}\label{sC}
f_D({\Delta t}/{{t_w }^{\beta}})=t_w^\alpha\overline{C}= \int
t_w^{-\alpha}P(C)(t_w^\alpha C) d(t_w^\alpha C).
\end{equation}
For a given ${\Delta t}/{{t_w }^{\beta}}$, $f_D$ is equal to a
constant, and Eq. (\ref{sC}) requires that $t_w^{-\alpha}P(C)$
should only depend on $t_w^\alpha C$. In other words,
$t_w^{-\alpha}P(C)$ is a function of ${\Delta t}/{{t_w }^{\beta}}$
and $t_w^\alpha C$. Then we define $F_D$ as:

\begin{equation}
F_D(t_w^\alpha C, \Delta t/t_w^\beta)=t_w^{-\alpha}P(C).
\end{equation}
A similar formula can be obtained for $P(\chi)$. Eventually, we
obtain the scaling ansatz of $P(C)$ and $P(\chi)$ shown in Fig.
\ref{pdf}:

\begin{equation}
\begin{array}{ll}
P(C)&=t_w^{\alpha}F_D(t_w^\alpha C, \Delta t/t_w^\beta), \\
P(\chi)&=t_w^{\alpha}F_M(t_w^\alpha \chi, \Delta t/t_w^\beta),
\end{array}
\end{equation}
where the universal functions $F_D(x,y)$ and $F_M(x,y)$ satisfy
\begin{equation}
\begin{array}{ll}
\int F_D(x,y)dx = f_D(y), \\
\int F_M(x,y)dx = f_M(y).
\end{array}
\end{equation}

\section{Study of the PDF of the autocorrelation function}
\label{power law}

The PDF of the autocorrelation function follows a modified power
law $t_w^{-\alpha}P(C)\propto(t_w^\alpha C+C_0)^{-\lambda}$, as we
see in Figs. \ref{pdf}a, \ref{pdf}b and \ref{pdf}c. From Eq.
(\ref{sC}), we have
\begin{equation}
\begin{split}
t_w^\alpha\overline{C}&= \int_{0}^{C_{cut}}
t_w^{-\alpha}P(C)(t_w^\alpha C) d(t_w^\alpha C) \\
&= \frac{\int_{0}^{C_{cut}} (x+C_0)^{-\lambda} x
dx}{\int_{0}^{C_{cut}} (x+C_0)^{-\lambda} dx}.
\end{split}
\end{equation}
The cutoff $C_{cut}$ ($C_{cut}\gg C_0$) is introduced to make the
integral converge and we always take $\lambda \geq 1$, then
\begin{equation}
t_w^\alpha\overline{C}=
\frac{C_0}{\lambda-2}[1-(\lambda-1)(\frac{C_{cut}}{C_0}+1)^{2-\lambda}].
\label {C0Ccut}\end{equation}

For $\lambda > 2$, the last term in Eq. (\ref{C0Ccut}) is
negligible and $t_w^\alpha\overline{C}\approx C_0/(\lambda-2)$
mainly depends on the short-time parameter $C_0$. For $\lambda < 2$,
we have $t_w^\alpha\overline{C}\approx
\frac{\lambda-1}{2-\lambda}C_0(\frac{C_{cut}}{C_0})^{2-\lambda}$
and the long-time parameter $C_{cut}$ dominates.

Following the previous discussion of $P(C)$, we define the $i$-th
tracer's local correlation function as:

\begin{equation}
C_{i}(t_w+\Delta t, t_w)=\langle\Delta x(t_w+\Delta
t,t_w)^2\rangle_i/2,
\end{equation}
where the average $\langle ... \rangle_i$ is calculated for only
the $i$-th tracer's trajectory by opening a small time window
$[\Delta t-t_{s},\Delta t+t_{s}]$ ($t_s \ll \Delta t$) at a given
$\Delta t$, and counting all the $C_{i}(t_w+\Delta t_j, t_w)$ with
$\Delta t_j\in[\Delta t-t_{s},\Delta t+t_{s}]$ into the statistics
of $P(C)$. We should note that if there is no average $\langle ...
\rangle_i$, $P(C)$ can be reduced to a simple form:

\begin{equation}
P(C)=P(\langle\Delta x^2\rangle_i/2)\xrightarrow{\langle ...
\rangle_i\rightarrow 0} P(\Delta x^2/2).
\end{equation}
This form can be further reduced since
\begin{equation}
P(\Delta x^2/2)d(\Delta x^2/2)=P(\Delta x)dx,
\end{equation}
therefore at the limit of the average $\langle ...
\rangle_i\rightarrow 0$,
\begin{equation}\label{Px}
P(C)=P(\langle\Delta x^2\rangle_i/2)\xrightarrow{\langle ...
\rangle_i\rightarrow 0} P(\Delta x^2/2)=P(\Delta x)/{\Delta x}.
\end{equation}

Therefore our $P(C)$ is related to the probability $P(\Delta x)$
usually studied in previous works \cite{Weeks_sci}. The reader
with enough endurance to have reached this point in the paper may
realize the apparent contradiction between our power-law scaling
and the typical stretched exponential behaviour found in other
studies of local fluctuations. In this regard we may first say
that previous work considered the supercooled regime while we work
in the glassy regime. Thus, both regimes may show different type
of scaling. When we analyze the scaling of $P(\Delta x)$ in our
glassy system we find that it can be approximately fitted by both
a broad tail power law behavior for large $\Delta x$, consistent
with equation \ref{mpl}, but also by a stretched exponential,
which is consistent with previous works \cite{Weeks_sci} for the
supercooled regime. On the hand $P(\langle\Delta x^2\rangle_i/2)$
can not be fitted by stretch exponential and it is only fitted by
the power-law scaling. Therefore, we conclude that the power
scaling should be the proper scaling for $P(\Delta x)$ and $P(C)$
in the glassy regime.

\section{Summary}

We have presented experimental results on the scaling behavior of
the fluctuations in an aging colloidal glass. The power-law
scaling found to describe the transport indicates the slow
relaxation of the system. A universal scaling form is found to
describe all the observables. That is, not only the global
averages, but also the local fluctuations. The scaling ansatz,
however, cannot be described under present models of spin-glasses,
but it is more akin to that observed in elastic manifolds in
random environments suggesting that our system may share the same
universality class.

\clearpage

\end{document}